\documentclass[conference]{IEEEtran}
\renewcommand{\baselinestretch}{1.02}

\ifCLASSINFOpdf
\else
\fi

\usepackage[dvips]{graphicx}
\usepackage{url}

\begin{document}
%
\title{CAN Disabler: Hardware-based Prevention method of Unauthorized Transmission in CAN and CAN-FD networks}

\author{
    \IEEEauthorblockN{
        Ryo Kurachi\IEEEauthorrefmark{1},
        T. David Pyun\IEEEauthorrefmark{2},
        Shinya Honda\IEEEauthorrefmark{1},
        Hiroaki Takada\IEEEauthorrefmark{1},
        Hiroshi Ueda\IEEEauthorrefmark{3},
        Satoshi Horihata\IEEEauthorrefmark{3}}\\
    \IEEEauthorblockA{
        \begin{tabular}{ccc}
            \begin{tabular}{@{}c@{}}
                \IEEEauthorrefmark{1}
                    Graduate School of Information Science,\\
                    Naogya University,\\
                    Japan\\
                    kurachi@nces.is.nagoya-u.ac.jp,\\ \{honda, hiro\}@ertl.jp
            \end{tabular} & \begin{tabular}{@{}c@{}}
                \IEEEauthorrefmark{2}
                    Lead Engineer,\\
                    Software Group,\\
                    Sumitomo Electric Wiring Systems, Inc.,\\
                    USA\\
                    TPyun@sewsus.com
            \end{tabular} & \begin{tabular}{@{}c@{}}
                \IEEEauthorrefmark{3}
                    AutoNetworks Technologies, Ltd., \\
                    Sumitomo Electric Industries, Ltd.,\\
                    Japan\\
                    \{ueda-hiroshi, horihata\}@sei.co.jp
            \end{tabular}
        \end{tabular}
    }
}

\maketitle

\begin{abstract}
There have been a quite number of security attack cases against Controller Area Network (CAN) reported in recent years, but in the meantime no security function is included in the ECU in the current in-vehicle control network.
Thus, in-vehicle control networks particularly require cost-effective security features.
In this paper, therefore, we propose a method to block unauthorized CAN-bus access using a CAN controller that is modified to prevent the ECU from transmitting messages to the CAN-bus at abnormal frequency.
Then, we have also demonstrated the effectiveness of the disabler on CAN with flexible data-rate (CAN-FD) buses.
\end{abstract}


%
\IEEEpeerreviewmaketitle

\section{Introduction}
There are more than 70 electronic control units (ECU) used in modern vehicles [1]. These ECUs are connected to such as Controller Area Network (CAN) [2], Local Interconnect Network (LIN) [3], and FlexRay [4] to perform its control function. CAN, in particular, is the most widely used protocol for the in-vehicle control network and is used in many vehicles sold currently in the market.

For the purpose of improving passenger comfort and communication services, systems that connect external networks, such as mobile telephone networks, and the in-vehicle control systems have been offered in recent years. Some of these services and systems particularly interlink mobile equipment such as smartphones and the in-vehicle control systems. It is increasingly important for vehicles to provide functionality by interlinking with various equipment and external systems.

Meanwhile, there have been quite a number of case reports on attacks against vulnerabilities or protocol weaknesses of external networks and equipment. It is a challenge to determine how to protect in-vehicle control systems that should be highly safe and is required to perform in real time.

\subsection{Motivation}
There have been quite a number of attack cases against in-vehicle control systems reported in recent years [5, 6, 7]. Koscher et al. demonstrate that rewriting a vehicle ECU program enables unauthorized CAN messages being transmitted. Furthermore, Valasek et al. demonstrate that unauthorized devices on the CAN-bus enable unauthorized transmission of messages [6]. A recent research by Miller et al. demonstrate that rewriting the program of an ECU connected to mobile phone networks enables unauthorized messages to be transmitted [7]. As shown in these sample cases, there is no end to attack cases enabled by forged CAN messages transmitted due to rewritten programs of ECUs that are connected to external systems.

Inclusion of Secure Boot in the ECUs as a countermeasure of these attacks has been discussed [8]. Secure Boot, however, is not sufficient enough to protect ECUs against possible attacks under situations where malware is installed after Secure Boot is enabled, combined with various usage environments such as constantly active ECUs and users that download and use various applications for on-board infotainment systems.

In our study, therefore, we propose a hardware-used device disabler for CAN messages. More concretely, this device disabler intends to prevent ECUs from being reprogrammed by using an unauthorized transmission monitoring feature that is installed in the improved CAN controller and to prevent unauthorized CAN messages to be transmitted from a malwareinfected ECU.

\subsection{The organization}
This paper is organized as follows. Sec. II is a brief overview of our subject, the in-vehicle control system, as well as an introduction to the ECU to which the proposed device disabler is applied. The proposed device disabler is explained in Sec. III and an implementation example is explained in Sec. IV. In Sec V, the evaluations and their results are given with a discussion given in Sec. VI. A summary and future development plans are given in Sec. VII.

\newpage

\section{Subject}
\subsection{In-vehicle electronic control systems}
There are many ways of using in-vehicle networks. In-vehicle network systems are used, for example, for collaborative control of the engine and the steering for the chassis systems and the transmission systems, for collaborative control of the engine and the motor on hybrid vehicles, and to control the keyless entry system, doors, the climate control, and meter indicators for the body systems. It is also used to link the car navigation system and audio equipment for the multimedia systems. Furthermore, in the parking assist system that helps the driver steer and park the car devices of respective sub-networks cross-functionally collaborate to perform assistance function. With the advancement in vehicle functions and technical advantages, the number of ECUs used in one vehicle has been increasing. The number of in-vehicle power lines and communication lines increases as the network scale becomes larger.

External interfaces such as the OBD-II port are standard equipment used for diagnostic function, and the network connection with external devices and networks is expected to be further developed along with the advancement of ITS. What is more, vehicles are more and more equipped with safety related systems such as the parking assist system and the brake assist system, which leads to possible risk of threat to these safety systems.

Fig.1 shows a schematic diagram of in-vehicle control system. The ECU that requires the device disabler proposed in this paper is applicable to ECUs connected to outside networks(V2X, ADAS, Car access module, Infotainment in Fig.1) and gateways that are required to be constantly active(Central Gateway in Fig.1). These ECUs and gateways are relatively larger in scale among other in-vehicle control systems, and it is hard to secure the reliability of software. The infotainment system that various applications are expected to be installed, in particular, has a problem of possible vulnerability due to multi-purpose OS being used. These ECUs, although vehicle manufacturers do not recommend in some cases, are also applicable to dongles (e.g. Mobile Devices C4oBD2 Dongle [9]) that are connected to OBD-II ports made by third party suppliers.

\begin{figure}[h]
   \begin{center}
      \scalebox{0.48}{
      \includegraphics[clip]{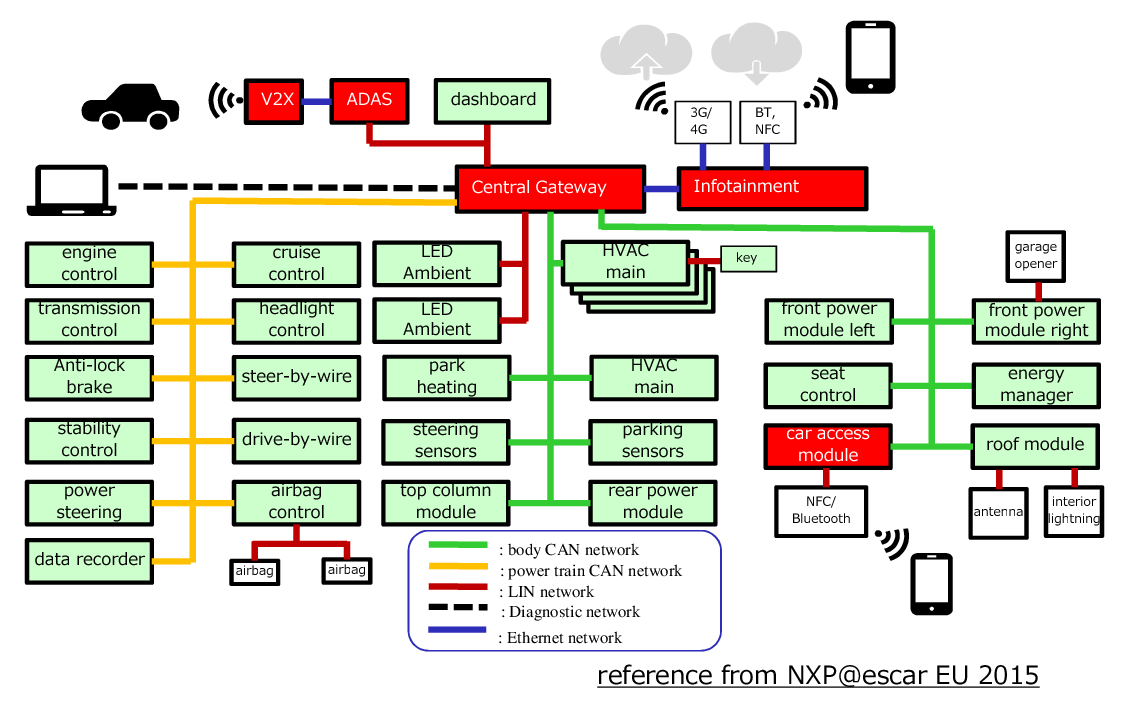}}
   \end{center}
\caption{ECU mapping in relation to subject in-vehicle control systems and the device disabler.}
\label{fig:sys}
\end{figure}

\subsection{Controller Area Network(CAN)}\label{sec:can}

CAN bus is a communication protocol standardized in ISO11898. It mainly specifies the 1st and 2nd layers of its OSI model with features as follows.

\begin{itemize}
\item {\bf Bus topology: } Bus topology is widely used with which multiple ECUs are connected to one communication line.
\item {\bf Multimaster: } When each node has a message to transmit, it can request a transmission and transmit the message to CAN-bus. Thus CAN messages and nodes can be easily added. 
\item {\bf Bus Arbitration: } If multiple nodes transmit a message to Can-bus at the same time, data competition occurs. To avoid this situation, the transmission right is adjusted using the CAN-ID. The highest priority CAN message is prioritized to be transmitted and low priority messages are delayed until the transmission of highest priority message is completed.
\item {\bf Mailbox: } A group of registers to transmit/receive messages is called a mailbox in the CAN communication controller. A typical CAN controller has multiple mailboxes for transmitting and receiving respectively (for example, there are 64 mailboxes for one CAN1 channel in V850E2/Px4 and each mailbox can be set as either outbox or inbox). The ECU often uses multiple outboxes or inboxes that are prepared by the communication controller, but not all of the mailboxes are always used.
\end{itemize}

\subsubsection{Threat}
Several existing automotive attacks are based on rewriting the software. In a modern automobile, most of the software and the data are stored in flash memory. In this systems, the security of the read/write access to flash memory is undoubtedly an issue of paramount importance. However, Kosher and Miller demonstrated an attack of rewriting software in real ECUs is possible under reasonable.

In addition, many of ECUs are designed by using a specialized operating system such as OSEK/VDX or AUTOSAR. However, these standards do not define the secure installation of software. Therefore, OEM needs to decide using several methods to secure installation and protection mechanisms such as signatures or obfuscation of program code.

Fig.2 illustrates an example of spoofed transmission on the CAN by a malicious ECU. Our proposed method would check the validity of the transmission events in each ECU to decrease the impact of this attack.

\begin{figure}[h]
   \begin{center}
      \scalebox{0.5}{
      \includegraphics[clip]{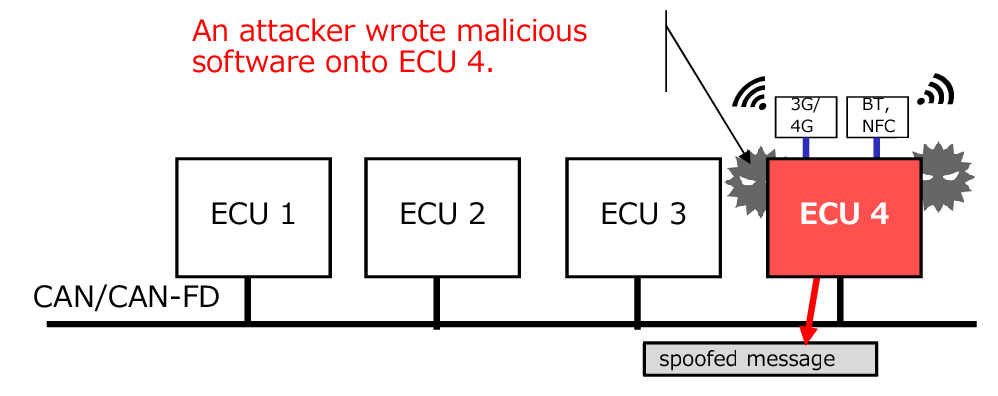}}
   \end{center}
\caption{Example of rewriting software of ECU.}
\label{fig:malicous}
\end{figure}

\subsubsection{Motivation and contributions}
We herein present the hardware-based prevention techniques that can protect in-vehicle networks from several types of CAN attacks

The main contributions of this work are as follows. To improve the security of in-vehicle systems, our proposed hardware-based techniques employs special security features, such as denial of service (DoS), unauthorized transmission detection by malicious software. To evaluate the performance on the CAN and CAN-FD environments, we conducted experiments on an implementation using the field programmable gate array (FPGA). In our system, the proposed device disabler requests security features only to check justification of transmiting messages by using the whitelist.
Because our proposed device disabler only can verify own transmiting request from its application software, the device disabler can not verify transmiting messages from other ECU.
Therefore, if possible, all needed ECUs must be equipped with this security functions. However, in terms of cost-efficiency, we consider the applicable level of the security feature depending on the types of ECUs.

In terms of security, our focus is on decreasing the effectiveness of malware and ECUs with compromised/malicious software, to prevent the above-mentioned attacks. Therefore, we do not address message authenticity and integrity because our proposal does not use cryptographic techniques.

\section{Proposed Method}\label{sec:prop}
In this section, the proposed method is explained, followed by demonstrations of the effectiveness of the proposed method against the existing security attack cases. 

\subsection{Proposed method: device disabler}
The purpose of this proposed method is to minimize possible influences of a program-rewritten ECU (hereinafter referred to as bastion ECU) on systems by setting an access restriction to CAN-bus for the bastion ECU, expecting that the program of an ECU connected to outside networks is rewritten by an attacker. Protection functions provided by the device disabler are as follows.

\begin{itemize}
\item {(Protection function 1) Usage restriction on used mailboxes:} The device disabler prohibits the use of unused mailboxes other than the ones that should be used by the bastion ECU to transmit messages.
\item {(Protection function 2) Whitelist restriction on sending messages:  } The device disabler prohibits transmitting messages other than CAN messages that should be transmitted by the bastion ECU.
\item {(Protection function 3) Transmission at abnormal frequency: } The device disabler controls the transmission frequency so that the bastion ECU does not transmit CAN messages more frequently than it should transmit.
\end{itemize}

Unauthorized messages can be easily transmitted from the bastion ECU once malware is installed because the current ECU is not equipped with these protection functions. Specifically the following threats exist.
\begin{itemize}
\item In the case of the protection function 1 being not equipped, malware spoofs the legitimate program of the bastion ECU and transmits unauthorized messages from unused mailboxes. 
\item In the case of the protection function 2 being not equipped, malware spoofs the legitimate program of the bastion ECU and transmits unauthorized CAN messages.
\item In the case of the protection function 3 being not equipped, malware transmits messages at abnormal frequency to easily perform DoS attacks against CAN-bus.
\end{itemize}

\subsection{Method to implement the proposed method}
The device disabler can be implemented by extending the hardware of the existing CAN controller. The device disabler is a hardware that intends to determine whether a CAN message is transmittable or not when its transmission is requested. The following evaluations 1 to 3 are performed using the evaluation procedure shown in Fig.2 to determine whether the requested message is authorized to be transmitted or not, assuming that information of transmittable CAN message (CAN-ID, DLC, etc.) and information such as its minimum transmission cycle and mailbox ID for transmission are set as the whitelist of the device disabler in the CAN controller.

\begin{itemize}
\item {\bf (Evaluation 1)} {The device disabler evaluates the mailbox whether it is authorized to be used for transmission. }
If a mailbox that is not authorized to be used for transmission is intended to be used, the device disabler discards the transmission. Meanwhile it keeps the transmission operation only if the mailbox is an authorized one, and it performs (Evaluation 2).
\item {\bf (Evaluation 2)} The device disabler evaluates, based on the whitelist, whether the CAN message is authorized to be transmitted or not.
If an unauthorized CAN message is to be transmitted, the device disabler discards this transmission meanwhile it keeps the transmission operation only if the message is an authorized one, and it performs (Evaluation 3). 
\item {\bf (Evaluation 3)} The device disabler evaluates the minimum transmission cycle of the requested transmission. 
If the transmission frequency of the requested transmission is higher than the preset minimum transmission cycle, the device disabler permits this transmission. If the transmission frequency is less than the minimum transmission cycle, the device disabler discards the transmission.
\end{itemize}

\subsection{Process to write in the whitelist}
In the proposed method, the method to write in the whitelist is important because this proposed method is invalid if the whitelist is falsified. Thus, the following two procedures are assumed to be followed when writing in the whitelist. In the case of the whitelist being written on runtime, it is required to prevent the whitelist from being rewritten once it is set. For this reason, it is assumed that if the following writing method 2 is used, the device disabler proposed in this paper is designed to set the whitelist during the period after the reset is released and till the rewriting completion register is set.

\begin{figure}[h]
   \begin{center}
      \scalebox{0.45}{
      \includegraphics[clip]{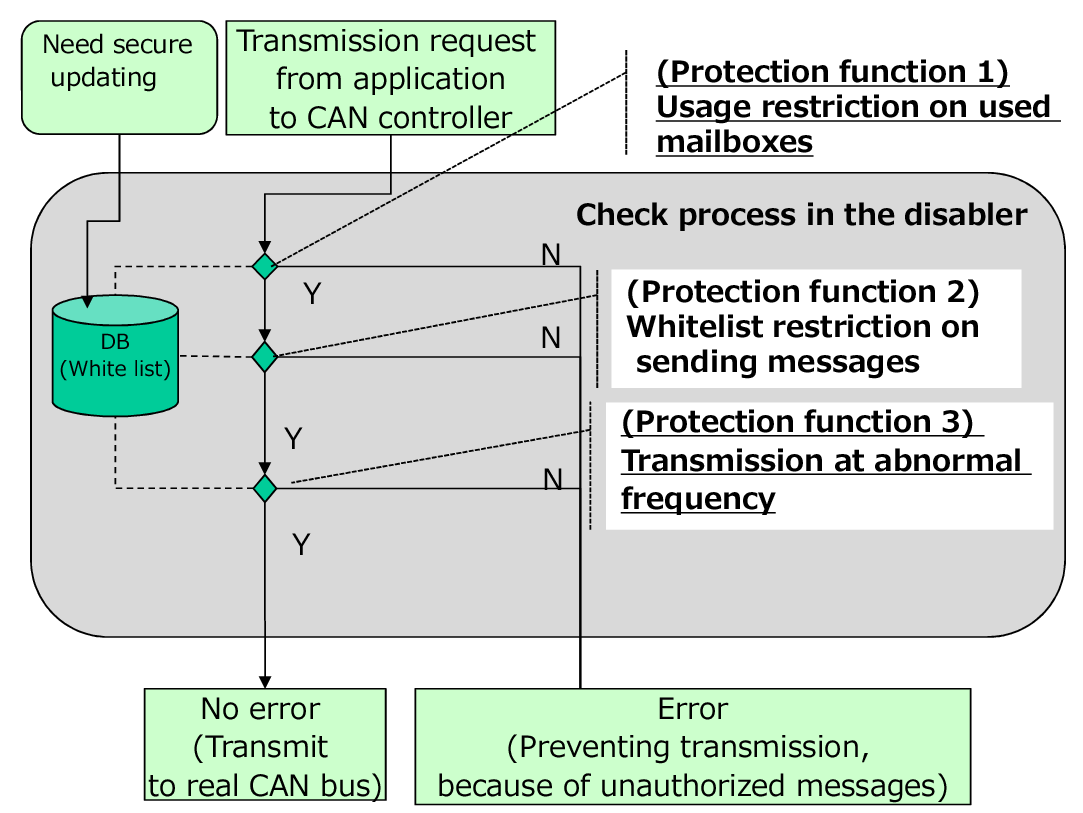}}
   \end{center}
\caption{Evaluation procedure for processing a transmission request}
\label{fig:check}
\end{figure}

\begin{itemize}
\item {\bf (Writing method 1)} The whitelist is written at delivery on a rewritable flash-ROM that is rewritable only with an authorized diagnosis tool. 
This type of method is used to set the clock supplied to the microcontroller with some of the current commercially available microcontrollers. A similar mechanism is assumed to be used to write in the whitelist.
\item {\bf (Writing method 2)} The whitelist is set in the device disabler using Secure Boot.
In this case, Secure Boot is required to be implemented using such as the secure element, and the hardware cost might increase compared with the existing ECU.
\end{itemize}

The writing method 1 is considered to have a particular affinity for the existing in-vehicle control systems and it is high in cost efficiency since the current existing ECUs do not implement Secure Boot.

\subsection{Assumed application case study}
We present one application case study that shows the type of vulnerability the proposed method is able to eliminate.

\subsubsection{Application case: protection against unauthorized transmission from external networks.}
OBD-II dongles, that are used to diagnose vehicle conditions from outside through the Internet networks for the purpose of using services such as metromile [10], are commercially available in the market. These dongles are attached for the purpose of discounting the insurance fee based on the driving mileage. Dongles are intended to receive mileage related information that is transmitted to the CAN network and upload the information to the server of service companies. There are, however, some security attack cases that an attacker exploits the vulnerability of a used OBD-II dongle, downloads malicious programs from the server of the attacker, and transmits unauthorized messages on CAN-bus. 
When this OBD-II dongle is equipped with the device disabler proposed in this paper, it is possible that unauthorized CAN messages are prevented from transmitting with the aforementioned (Evaluation 2) if the device disabler is set with information that there is no mailbox for transmission to CAN-bus.
In the case of falsified vehicle program presented by Miller et al. [7], it is possible to block the transmission of messages that are supposed to be transmitted from other ECUs even though messages transmitted from the program-rewritten bastion ECU are not blocked. 
According to these results, controlling message transmission with software, such as the device disabler proposed in this paper, is considered to be highly advantageous.

\section{Implementation}
We embedded the proposed device disabler onto an Altera FPGA board (DE0-NANO). We redesigned the existing CANFD controller IP to embed this device disabler. Fig.4 shows the developed system in the actual environment.

At first, we decided to implementat the whitelist. As a result, we added  registers for the whitelist as per the following tables:

\begin{table}[h]
\caption{Adding registers for the whitelist of the disabler.}
\label{tab:element}
\hbox to\hsize{\hfil
\scalebox{1.0}{
\begin{tabular}{l|l}
\hline\hline
register name                           & Objective                 \\ \hline
Disable mailbox information             & For protection function 1 \\
Sending message information             & For protection function 2 \\
Minimum intervals of sending messages   & For protection function 3 \\ \hline
\end{tabular}\hfil}
}
\end{table}

As shown in Table I, the "Disable mailbox information” register demonstrates the protection function 1. This register is implemented as a 1 bit register per each mailbox, and is used to enable or disable the corresponding mailbox. If the designer does not use several mailboxes, the user must set the disable condition of the corresponding mailboxes in advance. The ”Sending message information” and ”Minimum transmission cycles” are also to demonstrate the protection functions 2 and 3, respectively. These registers includes transmitting the CAN ID and corresponding to minimum time intervals.

Fig.5 shows the block diagram of the improved CAN-FD IP controller. The implemented CAN-FD controller consists of four parts: User register, Protocol Processing, Bit timing registers, and the counter process block. Our main extension is the counter prosess block and the protocol processor block to implement the algorithm for verify the tranmission request events presented in Fig.3. The counter process block is calculated for each transmission interval in the hardware. Additionally, User registers are almost the same for the existing CANFD IP, although several extensions were added for storing the whitelist information for each transmitting message, as shown in Table I.

\begin{figure}[h]
   \begin{center}
      \scalebox{0.8}{
      \includegraphics[clip]{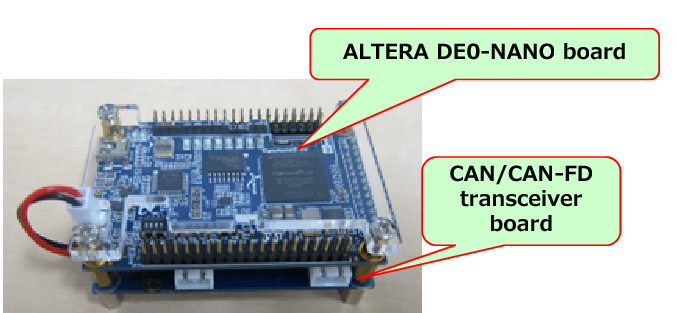}}
   \end{center}
\caption{Implementation environment of the disabler.}
\label{fig:environment}
\end{figure}

\begin{figure}[h]
   \begin{center}
      \scalebox{0.485}{
      \includegraphics[clip]{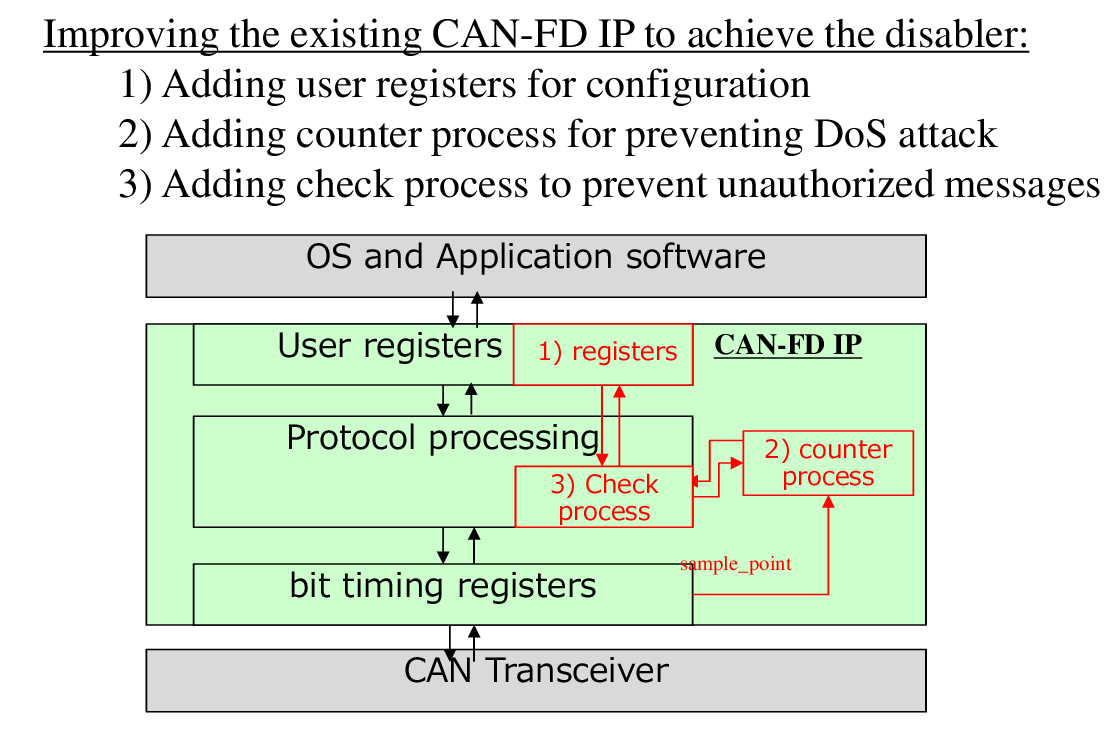}}
   \end{center}
\caption{Block diagram of the improved CAN-FD IP consisting of four blocks, including the three existing blocks (User register, Protocol Processing, Bit timing registers) and one added blocks (counter process).}
\label{fig:assess1}
\end{figure}

\subsection{Method to embed the minimum transmission cycle counter}
The embedding method of the counter (hereinafter called as minimum transmission cycle counter) is as follows. The minimum transmission cycle counter was embedded as a bit/time unit counter on the CAN network. The counter starts to count up after the initial transmission request is generated and to monitor the cycle. Thus the initial transmission request is not protected by the minimum transmission cycle counter. At the 2nd and following transmission requests, however, the counter is designed to discard transmission requests on the CAN controller without transmitting messages to CAN-bus.

\subsection{Method to embed the device disabler and its synthetic result}
The device disabler was embedded as a sub module in the existing CAN controller IP. The overall system consists of a DRAM controller and an on-chip RAM in addition to the NiosII soft core and a redesigned CAN controller. The system ’s synthetic results show that it consumes 23,888 logic elements. The CAN controller itself consumes 5374 logic elements. In this synthesis, there are 3106 more logic elements compared with the synthesis using the conventional CAN controller.

\section{Assessment}
We embedded the device disabler proposed in this paper in the FPGA. This assessment is conducted based on the premise that the writing method 1 described above is used.

\subsection{Assessment method}
The assessment was conducted for two subjects as follows. The first subject is the assessment of overhead in terms of the added device disabler. The device disabler is embedded in the hardware, but the standby time till the transmission start after a transmission request being processed increases due to the added inspection processing. In this paper, we demonstrate that this increased standby time is acceptable. The other subject is, for the purpose of showing the effectiveness of this proposed method, the assessment of the effect of the device disabler in case that the program is rewritten without authorization. More specifically we assess the device disabler to prove that it is effective if the program is rewritten.

\subsection{Assessment 1: Overhead of processing time}
The CAN controller IP, on which the device disabler proposed in this paper is embedded, needs more overhead required for filtering at the transmission of requests than the conventional CAN controller IP. Thus we measured the overhead process under the below conditions. A hardware counter was used in this measurement. The result shows that the processing time for the evaluation in Fig.2 is maximum 6.25~$\mu\mathrm{s}$ s after a transmission request is detected, and confirms that the delay time is within 3 to 4 bits at 500kbps CAN speed. These results confirm that the overhead of the CAN controller IP is in a suitable range since it is small enough compared with the transmission cycle of CAN messages in which time relating to the inspection process is cyclically transmitted, although it needs more overhead than the conventional CAN controller IP.

\subsection{Assessment 2: Effectivity of the device disabler}
In this section, the assessment is conducted for the situation that the program is rewritten without authorization.

\subsubsection{Assessment 2-1: Prevention of unauthorized CAN messages being transmitted from the ECU that the program is rewritten without authorization.}:
When a CAN controller with the device disabler embedded is used, only the CAN messages on the CAN ID registered in the whitelist are transmitted even if the program of ECU is rewritten. To confirm this behavior,  we created a program that requests the transmission of arbitrary CAN ID from the application and checked that only CAN messages registered in the whitelist were transmitted.

\begin{itemize}
\item{Case 1: Assesment results of protected function 1: Fig.6 shows the ECU never transmitting any messages when all transmitting mailboxes are disabled.}
\begin{figure}[h]
   \begin{center}
      \scalebox{0.445}{
      \includegraphics[clip]{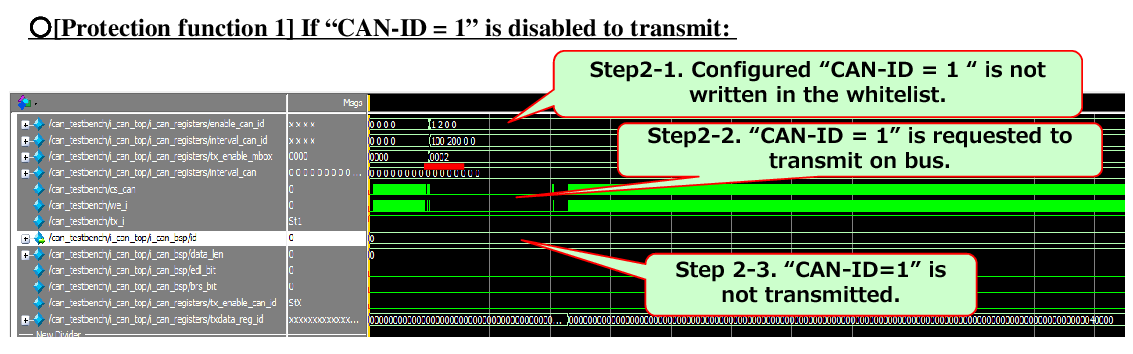}}
   \end{center}
\caption{Assesment results of protected function 1.}
\label{fig:assess1}
\end{figure}

\item{Case 2: Assesment results of protected function 2: Fig.7 presents the effectiveness of the disabler as caused by sending message filtering with a whitelist.}

\begin{figure}[h]
   \begin{center}
      \scalebox{0.45}{
      \includegraphics[clip]{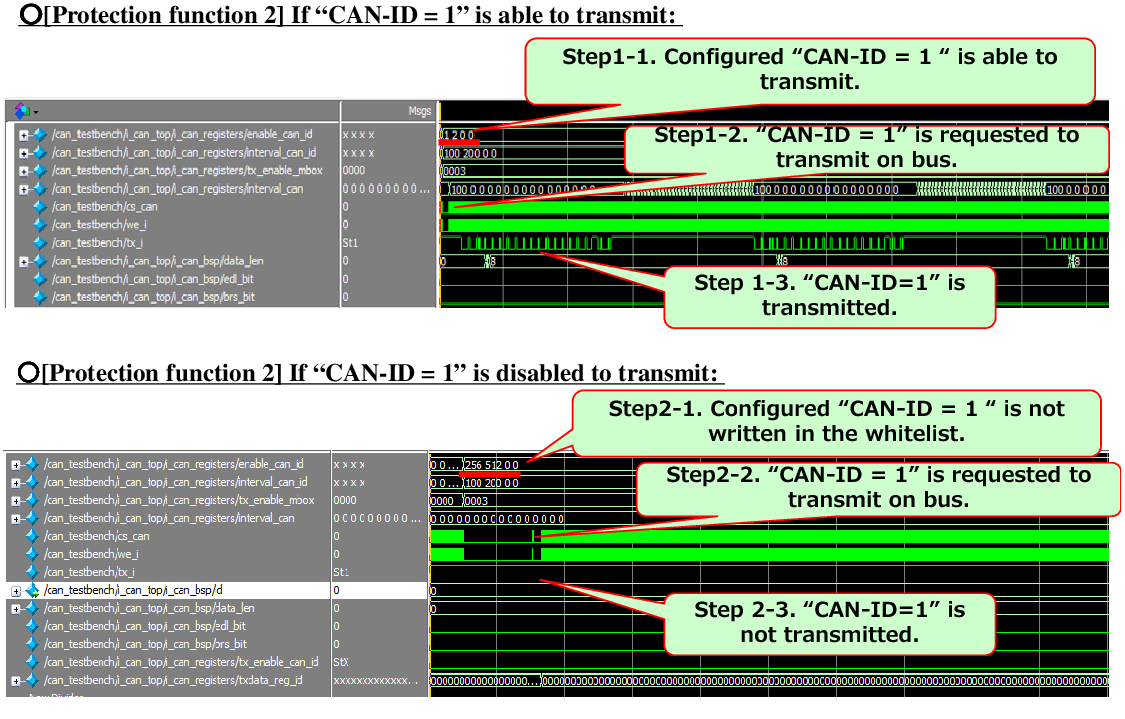}}
   \end{center}
\caption{Assesment results of protected function 2.}
\label{fig:assess2}
\end{figure}
\end{itemize}

\subsubsection{Assessment 2-2: Frequency control of message transmission on CAN networks from a bastion ECU that was rewritten without authorization:}
In the function provided by the proposed device disabler, it is possible to block denialof- service attacks (DoS attacks) from a bastion ECU where the program is rewritten without authorization, because the minimum transmission frequency is held in the whitelist. For a more concrete example, when the minimum transmission cycle of a CAN message is set to 2ms, only one CAN message is permitted to be transmitted during this 2ms period. In case there are multiple messages from this ECU however, the setting of minimum transmission cycle requires attention, because the frequency control depends on the number of messages and the set minimum transmission cycle may not be an effective measure against DoS attacks. In this assessment, if a program that requests a transmission approx every 10ms is created, but a whitelist to prevent false transmission over the CAN network is not employed, transmissions will continue over the CAN bus according to the requested frequency and intervals, so a Dos attack is possible. However, if device disabler is embedded on the network, its effective against DoS attacks was confirmed as there is an interval time check for the minimum transmission cycle.

\begin{figure}[h]
   \begin{center}
      \scalebox{0.48}{
      \includegraphics[clip]{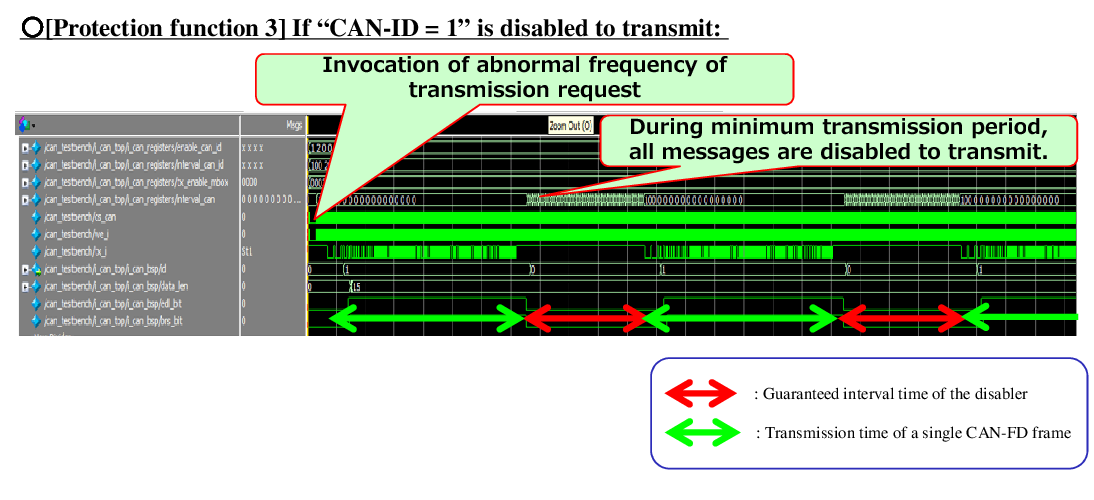}}
   \end{center}
\caption{Assesment results of protected function 3 by DoS attack from malware.}
\label{fig:assess3}
\end{figure}

\subsection{Discussion}
\subsubsection{Discussion 1: Useful range of the device disabler}
The device disabler proposed in this paper is not able to protect the ECU program from falsification. The device disabler, however, is able to prevent, under certain conditions, spoofed messages from being transmitted from the ECU whose program was rewritten without authorization. Thus, for best protection practices, it is desirable to implement a CAN controller with the device disabler embedded and Secure Boot. On the other hand, implementing Secure Boot in all the ECUs used in the in-vehicle control systems makes the cost extremely high as Secure Boot requires secure elements implemented in the existing ECUs. Overhead required for startup is also a concern. Therefore, it is desirable to implement both functions in ECUs that are connected to outside devices or devices that involve in in-vehicle control in particular, as shown in Chart 1. Either Secure Boot or the device disabler may be selected to apply to some safety-related systems of the in-vehicle control systems and the body systems. In this case, the secure element or the device disabler proposed in this paper should be selected to be implemented. This tradeoff condition and overhead for startup due to Secure Boot are future issues to be resolved.

\begin{table}[h]
\caption{Useful range of the device disabler.}
\label{tab:element}
\hbox to\hsize{\hfil
\scalebox{0.9}{
\begin{tabular}{l|cc|l}
\hline\hline
ECU Lv.             & Secure boot  &  The device disabler  & Type of ECUs                \\ \hline
Level 1             & Need         &  Need                 & GW, Engine, DCM, H/U        \\
Level 2             & No Need      &  Need                 & Safety-Critical system      \\
Level 3             & No Need      &  No Need              & None Safety-Critical system \\ \hline
\end{tabular}\hfil}
}
\end{table}

\subsubsection{Discussion 2: Comparison with the conventional researches.}
There have been researches on firewalls and IDSs that filter communications from outside in the past. Otsuka et al. [11] propose a control system that can be easily implemented using software. This control system monitors the message reception interval using an application on the software, focusing on the behavior of periodic message transmissions from respective CANs in the vehicle system. Brooks [12] announced that he developed an IDS system for periodically transmitted messages and event messages, but the details of this system are still not yet published. Muter et al. [13] propose an entropybased IDS system, but it is considered to be quite difficult to realize advanced arithmetic processing because ECUs in the in-vehicle control systems are equipped with low-speed and cost-effective CPUs. Miller et al. [6] also refer to a method to monitor transmission frequency. Sekiguchi et al. propose a hub operation using a whitelist but the update process of the whitelist hub is not yet clarified [14]. Furthermore, their proposed method is based on the hub configuration and not intended to be used with the ECU that has one CAN transmission port. Ujiie et al. propose a software-based filtering (firewall) of communication from external devices or networks by setting a static filter. Their proposed method mentions Secure Boot but no indication is made for the validity on the vehicle infotainment system that is potentially infected with malware. 

Furthermore, Herber et al. propose a method to minimize the possible influence of DoS attacks by time-sharing the access to CAN-bus [16, 17]. This method is in fact possible to localize the influence of DoS attacks due to malware infection, but it does not mention about unauthorized CAN messages. Based on these background reasons, we consider that our proposed method is effective for H/U and the navigation system that are equipped with multiple applications.

To evaluate the effectiveness of our proposed method, we compare it qualitatively with other methods. The evaluation criteria are: (1) Applicable to single core ECUs, (2) Preventng DoS Attack from malicous ECUs, (3) Disabling unused mailbox, (4) Secure updating method of the whitelist, (5) Preventing spoofed mesages. The results are shown in Table III.

According to Table III, in a case where the systems allows transmission without preventing spoofed messages, our proposal is a more secure solution than the other existing methods in single-core ECUs, but the reliability in multi-core ECUs is lower than TUM in paper [15]. However, we conclude that our proposal is reasonable for most of existing single-core ECUs, because the TUM method only effect multi-core ECUs which implement specific inter-core communication protocol such as AXI-bus.

\begin{table}[h]
\caption{Comparison of our proposal and existing methods.}
\label{tab:element}
\hbox to\hsize{\hfil
\scalebox{0.85}{
\begin{tabular}{l|ccc}
\hline\hline
                                         & Our proposal  &  Matsumoto   & TUM  \\ \hline
Applicable to single core ECUs           & yes           &  yes         & no   \\
Preventing DoS Attack from malicous ECUs & yes           &  yes         & yes  \\
Disabling unused message box             & yes           &  no          & no   \\
Preventing spoofed mesages               & no            &  no          & no   \\ \hline
\end{tabular}\hfil}
}
\end{table}

\section{Conclusion}
In this paper, we propose a hardware-based device disabler that intends to protect CAN networks by extending the hardware of the CAN controller in case of the ECU program being falsified. In this paper, we assessed the device disabler that was embedded in a FPGA. From the results it is confirmed that the device disabler is effective as a protector against the existing security attack cases. In addition, the results prove that the device disabler is capable of minimizing possible influence on existing CAN networks if it operates as a bastion. The future challenges include the improvement of filtering capability of the protector in case of spoofed CAN messages being sent by a bastion ECU and the comparison with Secure Boot.


\section*{Acknowledgment}

This work was supported by Ministry of Internal Affairs and Communications (MIC) in Japan, Strategic Information and Communications R\&D Promotion Programme (SCOPE) Grant Numbers 152106005.
This work was supported by JSPS KAKENHI Grant Number 16K16025.



%

\end{document}